\documentclass[final,3p,times,twocolumn]{elsarticle}
\usepackage{graphicx}
\usepackage{hyperref}
\usepackage{subfigure}
\usepackage{amsmath}
\usepackage{bm}
\journal{Computer Physics Communications}
\begin{document}

\begin{frontmatter}

\title{ ATLAS: A Real-Space Finite-Difference Implementation of Orbital-Free Density Functional Theory}
\author[fad]{Wenhui Mi\fnref{ft1}}
\author[fad]{Xuecheng Shao\fnref{ft1}}
\author[fad]{Chuanxun Su}
\author [fad]{Yuanyuan Zhou}
\author [fad]{Shoutao Zhang}
\author [fad]{Quan Li}
\author [fad]{Hui Wang}
\author [sad]{Lijun Zhang}
\author [tad]{Maosheng Miao}
\author[fad]{Yanchao Wang\corref{cor}}
\ead{wyc@calypso.cn}
\author [fad]{Yanming Ma\corref{cor}}
\ead{mym@calypso.cn}
\address [fad]{State Key Laboratory of Superhard Materials, Jilin University, Changchun 130012,China}
\address [sad]{College of Materials Science and Engineering, Jilin University, Changchun 130012,China}
\address [tad]{Department of Chemistry and Biochemistry, California State University Northridge, 18111 Nordhoff St., Northridge, CA 91330, USA}
\cortext [cor]{Corresponding author }
\fntext[ft1]{These two authors contributed equally.}

\begin{abstract}
Orbital-free density functional theory (OF-DFT) is a promising method for large-scale quantum mechanics simulation as it provides a good balance of accuracy and computational cost. Its applicability to large-scale simulations has been aided by progress in constructing kinetic energy functionals and local pseudopotentials. However, the widespread adoption of OF-DFT requires further improvement in its efficiency and robustly implemented software. Here we develop a real-space finite-difference method for the numerical solution of OF-DFT in periodic systems. Instead of the traditional self-consistent method, a powerful scheme for energy minimization is introduced to solve the Euler--Lagrange equation. Our approach engages both the real-space finite-difference method and a direct energy-minimization scheme for the OF-DFT calculations. The method is coded into the ATLAS software package and benchmarked using periodic systems of solid Mg, Al, and Al$_{3}$Mg. The test results show that our implementation can achieve high accuracy, efficiency, and numerical stability for large-scale simulations. 
\end{abstract}
\end{frontmatter}
\section{\label{introduction}Introduction}
Computational simulation is a powerful tool for predicting material properties and understanding the physics underlying experimental observations.\cite{carter2008challenges} Reliable simulation relies on advanced computational theories and methods, and in recent decades many efficient approaches with different levels of accuracy have emerged to receive remarkable success; e.g., quantum-mechanical\cite{PhysRev.136.B864,PhysRev.140.A1133,goedecker1999linear} and empirical potential methods.\cite{PhysRevB.29.6443,PhysRevLett.50.1285} 

Quantum mechanical approaches based on the Kohn--Sham (KS) density functional theory (DFT) \cite{PhysRev.136.B864,PhysRev.140.A1133} allow accurate descriptions of materials' properties, but are computationally demanding. They require evaluation of the kinetic energy term related to the computation of single-electron wave-functions. The calculation of electron density needs to consider 3$N_{e}$ degrees of freedom, and thus the computational cost scales as a cubic relation of $O(N_{e}^3)$,\cite{RevModPhys.64.1045} where $N_{e}$ is the total number of electrons of the system. The large computational cost limits KS-DFT to only small systems with unit cells of up to only hundreds or thousands of atoms,\cite{RevModPhys.64.1045} which makes its traditional implementation unsuitable for studying complex systems (e.g., surfaces, interfaces, nanomaterials, and biomaterials) or macro-scale features (e.g., grain boundaries, dislocations, and cracks) that require large-scale simulations using cells of tens of thousands or millions of atoms.\cite{Hung2009163,carter2008challenges}

 Over the past two decades, many linear scaling techniques have been developed in an effort to reduce the cubic scaling of traditional KS-DFT.\cite{goedecker1999linear,bowler2012mathcal} However, they depend on both the ``nearsightedness''  principle\cite{goedecker1999linear,bowler2012mathcal} and the concept of ``locality'',\cite{PhysRevLett.76.3168} and therefore scale linearly only for systems containing a large number of atoms. There is an unavoidable crossover between cubic and linear scaling.\cite{carter2008challenges,goedecker1999linear,bowler2012mathcal} Moreover, linear scaling requires a band-gap structure or localized electronic structure, and appears not to function for metals.\cite{carter2008challenges,Hung2009163}

Alternative approaches related to parameter fitting have therefore been designed using empirical interatomic potentials.\cite{PhysRevLett.50.1285,PhysRevB.29.6443} These simulations require much less computational cost and are computationally capable of dealing with macro-scale problems, but they suffer notable shortcomings in terms of accuracy and transferability.\cite{carter2008challenges,Hung2009163} More severely, these simulations completely neglect the properties of electrons, which are fundamentally important to various aspects of chemistry and physics. There is therefore an urgent need for a reliable quantum-mechanics-based method able to perform large-scale simulations.

Orbital-free (OF) DFT,\cite{karasiev2014progress,wesolowski2013,karasiev2012issues,karasiev2009,wang2002} is potentially an efficient theory for large-scale quantum mechanical simulations. The total energy within the OF-DFT scheme is expressed as an explicit functional of electron density in the Hohenberg--Kohn theorem,\cite{PhysRev.136.B864} and there is no need to deal with wave functions. Here the electron density, a simple function with three degrees of freedom, can uniquely determine the ground-state properties of a many-electron system. As such, the computational cost of OF-DFT scales quasilinearly with the number of atoms in the system, providing substantial advantages in numerical simplicity and efficiency for large-scale simulations. \cite{karasiev2014progress,wesolowski2013,karasiev2012issues,karasiev2009,wang2002}

The drawbacks of OF-DFT include two challenges to its realistic treatment of the kinetic energy density functional (KEDF) and ion--electron interactions.\cite{karasiev2012issues} First, the kinetic energy term is a sole functional of the electron density function; its construction determines the accuracy of OF-DFT, and significant progress has been made in the last two decades (Refs. [\cite{PhysRevB.45.13196,pearson1993ab,perrot1994hydrogen,smargiassi1994orbital,foley1996further,jesson1997thermal,PhysRevB.60.16350,zhou2005improving,PhysRevB.81.045206,xia2012can,PhysRevB.85.045126,shin2014enhanced}]) with the proposal of various encouraging KEDFs. These functionals have been successfully tested in many systems, including nearly free-electron-like systems and semi-conductors.\cite{xia2014orbital,PhysRevLett.112.145007,danel2015equation} Second, the lack of particular orbitals leads the ion--electron interaction to be described only by the local pseudopotentials (LPPs), and previous studies have sought to construct various LPPs. Empirical\cite{wang2002,karasiev2009,wesolowski2013,PhysRevB.51.14001,PhysRevB.7.1295} and ``bulk derived'' LPPs\cite{PhysRevB.69.125109} have been developed and successfully applied to various metals and semiconductors,\cite{Hung2009163,xia2014orbital} but a notable challenge is the construction of a good LPP from an existing non-LPP without appeal to any bulk- or aggregate-system KS calculations.\cite{karasiev2012issues} We recently developed an optimized effective potential (OEPP) scheme to construct full first-principles LPPs from existing non-LPPs\cite{detail} to enhance the transferability of the pseudopotential. Our OEPP worked well for a large number of elements, and the transferability of the LPP was found to be an intrinsic property of elements.

The ground-state energy $E_{min}$ and electron density $\rho$ in OF-DFT\cite{wang2002,karasiev2012issues} can be obtained by minimizing the total energy $E[\rho]$ of the system with respect to the trial electron density $\rho$. The following minimization equation is non-linear and multidimensional:\cite{PROFESS1.0}
\begin{equation}
E_{min}=\underset{\rho}{min}\left\{E[\rho]-\mu(\int_{\Omega}\rho(\bm{r})d\bm{r}-N_{e});\rho \ge0 \right\},
\label{eq:mini}
\end{equation}
where $\mu$ is a Lagrange multiplier used to enforce the constraint that the total number of electrons $N_{e}$ is conserved, and $\Omega$ is the whole space for the simulation.

Currently, there are two main procedures for solving Eq. (\ref{eq:mini}).\cite{karasiev2012issues} The first is to seek a direct solution of the equation by minimizing the total energy with respect to the electron density $\rho$.\cite{garcia2007efficient,hung2012preconditioners,PROFESS1.0,PhysRevB.78.045105} The well-known PROFESS code was built this way,\cite{PROFESS1.0,PROFESS2.0,Chen2015228} and has been successfully used to investigate many large-scale problems.\cite{ho2009mechanical,shin2012orbital,chen2013melting} The other procedure is to transform Eq. (\ref{eq:mini}) into an ordinary KS-like equation that can be solved self-consistently by any KS computer program.\cite{levy1984exact,deb1983new,levy1988exact} However, recent research\cite{karasiev2012issues,chan2001thomas,parr1989density} has shown that the iterative self-consistent procedure for OF-DFT does not work properly for large systems. Moreover, the non-convergence problem is not solved, and the underlying reason for this remains unclear.\cite{karasiev2012issues}

In this work, a real-space finite-difference method for solving the OF-DFT Euler--Lagrange equation (Eq. (\ref{eq:mini})) for periodic systems is developed by direct minimization. As shown previously, a real-space finite-difference method provides three obvious advantages:\cite{PhysRevB.69.075101,schofield2012spectrum,RevModPhys.72.1041,hirose2005first,} (i) the method is independent of any basis, simplifying its implementation;\cite{chelikowsky1994finite} (ii) a real-space method is advantageous for large-scale parallel calculations due to its avoidance of the fast Fourier transform (FFT) method for the reciprocal-space approach and the serious drawback in the need for ``all-to-all communication''\cite{bowler2012mathcal} during parallel calculation; and (iii) there is no barrier to switching between a periodic and a non-periodic system for a real-space approach. Particularly, we find that the finite-difference method is computationally more efficient in dealing with the Laplace, gradient, and divergence operators than the FFT-based method.

Our method is coded into \textit{Ab initio} orbiTaL-free density functionAl theory Software (ATLAS) and is benchmarked in periodic systems of Mg, Al, and Al$_{3}$Mg. Our current implementation of OF-DFT is shown to be numerically accurate, stable, and efficient.

The remainder of this paper is organized as follows. Section II gives the theory. The OF-DFT differential equation is presented for illustration,  followed by detailed real-space representations of the finite-difference method and the direct energy minimization method to obtain the ground-state electron density. Section III reports testing results on Mg, Al, and Al$_{3}$Mg crystals to demonstrate the computational accuracy, efficiency, and stability of the procedure. Finally, conclusions are presented in Section IV.
\section{\label{Theory}Theory and Background}
\subsection{\label{of} OF-DFT Theory}
In the OF-DFT scheme,\cite{wang2002,karasiev2012issues} the ground state total energy of an $N$-electron system in a local external potential $V_{ext}(\bm{r})$ is a functional of electron density $\rho(\bm{r})$:
\begin{equation}
\label{conserving}
\int_{\Omega} \rho(\bm{r}) d\bm{r}  = N_{e}.
\end{equation}
The total energy functional $E[\rho]$ (atomic units, a.u., are used throughout the paper) can be written as follows:\cite{wang2002,karasiev2012issues} 
\begin{align}
E^{OF}[\rho]=&T[\rho]+E_{H}[\rho]+ E_{XC}[\rho]+\int V_{ext}(\bm{r})\rho(\bm{r}) d\bm{r}\notag \\
&+E_{i-i}(\{\bm{R}_{i}\}),
\label{energy}
\end{align}
where $E_{H}[\rho]$ is the Hartree electron--electron repulsion energy, $E_{XC}[\rho]$ the exchange--correlation energy, and $V_{ext}(\bm{r})$ the external potential representing the ion--electron interaction as given by local pseudopotentials in the OF-DFT scheme. $E_{i-i}$ is the interaction energy between ions, which is dealt with in our implementation using Ewald summation.\cite{Eward,Toukmaji199673,Essmann} The kinetic energy of the non-interacting electrons ($T[\rho]$) is an explicit functional of electron density. These KEDFs can be roughly categorized into two general types: local/semilocal and nonlocal. The former naturally scale linearly
 with system size, while the later scales quadratically owing to its double integral.\cite{PhysRevB.91.045124} Our program implements both local/semilocal KEDFs\cite{garcia2007kinetic,jctc2009} and the nonlocal Wang-Govind-Carter (WGC) KEDF\cite{PhysRevB.60.16350}. The WGC KEDF can be written as follows:
\begin{equation}
T^{WGC}[\rho] = T_{TF}[\rho]+T_{vW}[\rho]+T_{nl}[\rho],
\end{equation}
where $T_{TF}[\rho]$ is the Thomas--Fermi term assuming the limit of a uniform electron gas. It takes the form
\begin{equation}
T_{TF}[\rho]= C_{TF}\int_{\Omega}\rho(\bm{r}) d\bm{r},
\end{equation}  
where $C_{TF}=\frac{3}{10}(3\pi^{2})^{2/3}$. $T_{vW}[\rho]$ is the von Weizs\"acker (vW) term designed for a single-orbital system:
\begin{equation}
T_{vW}[\rho] = \int_{\Omega} \sqrt{\rho(\bm{r})}\left(-\frac{1}{2}\nabla^{2}\right)\sqrt{\rho(\bm{r})} d\bm{r}.
\end{equation} 
The third non-local term has the form
\begin{equation}
T_{nl}[\rho]=\int \rho^{\alpha}(\bm{r})k[\rho(\bm{r}),\rho(\bm {r}),\bm{r},\bm{r}']\rho^{\beta}(\bm{r}') d\bm{r}d\bm{r}'.
\end{equation}
This non-local term $T_{nl}[\rho]$ is Taylor expanded to achieve quasilinear scaling with system size via FFT.\cite{PhysRevB.60.16350,PhysRevB.78.045105}

Our approach decomposes the KEDFs into the $\lambda T_{vW}$ term and a remaining term $T_{\theta}[\rho]$ (which, when $\lambda= 1$, is the Pauli term\cite{karasiev2012issues}):
\begin{equation}
T[\rho]=\lambda T_{vW}[\rho]+T_{\theta}[\rho].
\end{equation} 
The vW kinetic potential term can be evaluated as follows: 
\begin{equation}
\label{eq:vW}
 \frac{\delta T_{vW}[\rho]}{\delta \rho}=\frac{1}{\sqrt{\rho}}(-\frac{\nabla^{2}}{2})\sqrt{\rho}.
\end{equation}
The finite-difference method is adopted here to obtain the vW kinetic potential term instead of FFT, as implemented in PROFESS.\cite{PROFESS1.0} 

Instead of minimizing $E^{OF}$ directly over the electron density, we rewrite the total energy as the functional of $\phi=\sqrt{\rho}$. Taking the constraint of Eq. (\ref{conserving}) into account by Lagrange's multiplier method, we define
\begin{equation}
L[\phi] = E^{OF}[\phi] -\mu \left\{\int_{\Omega}\phi^{2}(\bm{r})d\bm{r}-N\right\},
\end{equation}
then the gradient of $L$ with respect to $\phi$ is 
\begin{align}
\frac{\delta L[\phi]}{\delta \phi}&=\frac{\delta E^{OF}[\rho]}{\delta \rho}\frac{\delta \rho}{\delta \phi}-2 \mu\phi \nonumber \\
&=2\left\{\frac{\delta E[\rho]}{\delta \rho}-\mu\right\}\phi \nonumber\\
&=2\left\{H\phi-\mu \phi\right\},
\label{eq:gl}
\end{align}
where 
\begin{equation}
\label{eq:Hphi}
H\phi=-\frac{\lambda}{2}\nabla^{2}\phi(\bm{r})+V_{eff}(\bm{r})\phi(\bm{r}).
\end{equation}
 This equality is derived from Eq. (\ref{eq:vW}) and \begin{align}
\label{eq:veff}
V_{eff}(\bm{r})&=\frac{\delta T_{\theta}[\rho]}{\delta \rho}+\frac{\delta E_{H}[\rho]}{\delta \rho}+\frac{\delta E_{XC}[\rho]}{\delta \rho}
+V_{ion}(\bm{r})\nonumber \\
&=V_{\theta}(\bm{r})+V_{H}(\bm{r})+V_{XC}(\bm{r})+V_{ion}(\bm{r}).
\end{align}
The variational principle requires $\delta L/\delta \phi =0$, which leads to the Euler--Lagrange equation:
\begin{equation}
H\phi=\mu\phi.
\label{eq:el}
\end{equation}
This is a Schr\"odinger-like equation,\cite{deb1983new,levy1984exact,levy1988exact,jiang2004conjugate,karasiev2012issues} but much simplified as there is only one ``orbital'' with the minimal eigenvalue. This equation can then be resolved by minimizing the OF-DFT total energy with respect to $\sqrt{\rho}$.

\subsection{\label{differs} Real-Space Representations}
Real-space calculations are performed on grids, in which the values of the electron density distribution and effective potential are given on discrete Cartesian grid points. The real-space finite-difference expansion transforms the kinetic energy operator into a spare matrix, which has nonzero elements only in the vicinity of the leading diagonal.\cite{RevModPhys.72.1041,hirose2005first} The general form of the Laplacian with a Cartesian grid can be expressed as follows:
\begin{align}
\nabla^{2}\phi(x_{i},y_{j},z_{k})=&\sum_{n=-N}^{N}C_{n}\phi(x_{i}+nh_{x},y_{j},z_{k})\nonumber \\
&\sum_{n=-N}^{N}C_{n}\phi(x_{i},y_{j}+nh_{y},z_{k})\nonumber \\
&\sum_{n=-N}^{N}C_{n}\phi(x_{i},y_{j},z_{k}+nh_{z}),
\end{align}
where $N$ is the order of the finite difference expansion; $h_{x}$, $h_{y}$, and $h_{z}$ are the grid spacings in the $x$, $y$, and $z$ directions, respectively; and the $C_{n}$ coefficients are available in Ref.\cite{RevModPhys.72.1041,fornberg1988generation}. However, the Cartesian grid is incompatible with the periodicity of a non-orthorhombic unit cell. A new high-order finite-difference method for a non-orthorhombic grid has been proposed and successfully applied to periodic systems.\cite{PhysRevB.78.075109} We adopt it here. The general form of the Laplacian operator for a non-orthorhombic grid is as follows: 
\begin{equation}
\nabla^{2}=\sum_{i=1}^{6}f_{i}\frac{\partial^{2}}{\partial v_{i}^{2}}.
\end{equation}
We represent the Laplacian by a combination of derivatives along six nearest-neighboring $v_{i}$ directions: three original $\bm a_{i}$ (i=1,2,3) directions and three additional nearest-neighboring directions, where $a_{i}$ are the lattice vectors in real space. For the $f_{i}$ coefficient, refer to Ref. \cite{PhysRevB.78.075109} Note that the $H$ matrix (Eq. (\ref{eq:el})) is a spare matrix whose nonzero elements are confined within a diagonal band, and the extent of the nonzero elements in off-diagonal positions depends on the order of the finite difference expansion.
 \begin{figure}[!htbp]
 \subfigure[]{
    \label{fig:pp_Mg} 
    \includegraphics[width=8.0cm]{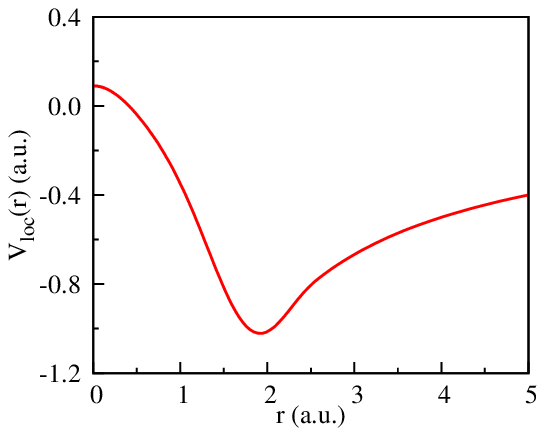}}
  \subfigure[]{
    \label{fig:pp_Al} 
   \includegraphics[width=8.0cm]{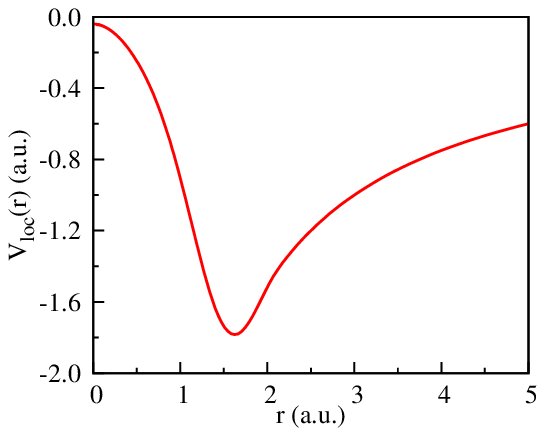}}
  \caption{Local optimized effective pseudopotential of Mg (a) and Al (b) in real space.}
\end{figure}The Hartree potential is determined by solving the Poisson equation:
\begin{equation}
\nabla^{2}V_{H}(\bm r) = -4\pi[\rho(r)-\rho_{0}(r)],
\end{equation}  
where $\rho_{0}(\bm r)$ is the average electron density of the system. For infinite periodic systems, we encountered the divergent problems on the ion--electron, ion--ion, and electron--electron interaction energies arising from the long-range Coulomb interaction $-Z/r$. Fortunately, the divergent problem converts into the singularity problem at $g = 0$ in reciprocal space. For a charge-neutral system, the singularity for Hartree electron--electron potential can be exactly canceled by adding up singularities encountered in the electron--ion and ion--ion potentials, and can therefore be neglected.\cite{martin2004electronic,PROFESS1.0} The Hartree potential $V_{H}(\bm{r})$ can thus be obtained as follows:
\begin{equation}
V_{H}(r)=FFT^{'} (\frac{4\pi}{|\bm{G}|^{2}}\rho(G)) \qquad (\rho(G=0)= 0),
\end{equation}
where $\rho(G)$ is the electron density in reciprocal space, and $FFT^{'}$ is a reverse FFT transform.
The ionic term $V_{ion}(\bm r)$ in Eq. (\ref{eq:veff}) can be constructed from $V_{loc}(\bm{r})$ (i.e., LPPs). We use our developed OEPP for LPPs. OEPPs for both Mg and Al are shown in Fig. \ref{fig:pp_Mg} and \ref{fig:pp_Al}, respectively. The theory of constructing OEPP is presented elsewhere.\cite{detail} For a periodic system, the ionic potential receives contributions from an infinite number of atoms, leading to a divergent summation of the long-range Coulomb term. To seek a solution, as mentioned above, the pseudopotential $V_{ion}(\bm{r})$ is then expressed in reciprocal space as follows:\cite{RevModPhys.64.1045,martin2004electronic}
\begin{eqnarray}
V_{ion}(\bm{G})\equiv&&\frac{1}{\Omega_{cell}}\int_{\Omega_{cell}}V_{ion}(\bm{r})exp(i\bm{G}\cdot \bm{r})d\bm{r}\nonumber \\
=&&\frac{1}{\Omega_{cell}}\sum_{\kappa=1}^{n_{type}}S^{\kappa}(\bm G)V_{loc}^{\kappa}(\bm G),
\end{eqnarray}
where $n_{type}$ is the number of atomic species, and for each atomic species $\kappa$ there are $n^{\kappa}$ identical atoms at positions $\bm{\tau}_{\kappa,j}$,$j=1,n^{\kappa}$, and $\Omega_{cell}$ is the unit cell volume. The structure factor $S(\bm{G})$ for each atomic species $\kappa$ is\cite{martin2004electronic}
\begin{equation}
S^{\kappa}(\bm G) =\sum_{j=1}^{n^{\kappa}}exp(i\bm{G}\cdot \tau_{\kappa,j}),
\end{equation} \\
and the form factor $V_{loc}(\bm{G})$ is\cite{martin2004electronic}
\begin{equation}
V_{loc}^{\kappa}(\bm G) =\int_{all  space}V_{loc}^{\kappa}(\bm r)exp(i\bm{G}\cdot \bm{r})d\bm{r}.
\end{equation}
The spherical symmetry of LPP allows the 3D Fourier transform to be represented by a 1D radial Fourier transform:\cite{martin2004electronic,PROFESS1.0}
\begin{align}
V_{loc}^{\kappa}(g)&=4\pi\int_{0}^{\infty}V_{loc}^{\kappa}(r)r^{2}\frac{sin(gr)}{gr}dr
\nonumber\\
&=V_{nc}^{\kappa}(g)-\frac{4\pi Z}{g^{2}}  \qquad   ( g\ne 0),
\end{align}

where the non-Coulomb potential $V^{\kappa}_{nc}(\bm{g})$ in reciprocal space can be written as
\begin{align}
V_{nc}^{\kappa}(g)=4\pi\int_{0}^{r_{cut}}(V_{loc}^{\kappa}(r)+\frac{Z}{r})r^{2}\frac{sin(gr)}{gr}dr.
\end{align}
At $g=0$, the Coulomb interaction is canceled as described above, leading to
\begin{equation}
V_{loc}^{\kappa}(\bm{g}=0)=4\pi\int_{0}^{r_{cut}}(r^{2}V_{loc}^{\kappa}(r)+Zr) dr,
\end{equation}
where $r_{cut}$ is the cutoff of core radii. In our implementation, $V_{loc}(r)$ is equal to $-Z/r$ when $r\ge r_{cut}$.

For a given grid spacing $h$, the size of the grid points can be determined as $\prod_{i=1}^{3}N_{i} $, where
\begin{equation}
N_{i}=\frac{|\bm{a}_{i}|}{h}.
\end{equation} 

For a given structure, the wave vector $\bm G$ is determined by
\begin{equation}
\bm{G}(n_{1},n_{2},n_{3})=n_{1}\bm{b}_{1}+n_{2}\bm{b}_{2}+n_{3}\bm{b}_{3},
\end{equation}
where $\bm{b}_{i}$ ($i=1, 2, 3$) are the primitive vectors in reciprocal space, and $n_{i}=(0,1,2\cdots N_{i})$ are integers.

Finally, the real-space local ion--electron pseudopotential $V_{ion}(\bm r)$ can be calculated by an FFT:
\begin{equation}
V_{ion}(\bm r)=FFT(V_{ion}(\bm{G})).
\end{equation}

As mentioned above, all the physical quantities in the real-space finite-difference formalism can be directly represented on discretized grid points with a uniform interval.\cite{PhysRevB.78.075109} Finally, Eq. (\ref{eq:Hphi}) can be expressed as the following discretized expression: 
\begin{align}
\label{eq:express}
&H\phi(u_{i},v_{j},w_{k})=\nonumber \\
-\frac{\lambda}{2}&[\sum_{n_{1}=-N}^{N}C_{n_{1}}\phi(u_{i}+\xi_{11}n_{1}h,v_{j}+\xi_{12}n_{1}h,w_{k}+\xi_{13}n_{1}h)\nonumber \\
+&\sum_{n_{2}=-N}^{N}C_{n_{2}}\phi(u_{i}+\xi_{21}n_{2}h,v_{j}+\xi_{22}n_{2}h,w_{k}+\xi_{23}n_{2}h)\nonumber \\
+&\sum_{n_{3}=-N}^{N}C_{n_{3}}\phi(u_{i}+\xi_{31}n_{3}h,v_{j}+\xi_{32}n_{3}h,w_{k}+\xi_{33}n_{3}h)\nonumber \\
+&\sum_{n_{4}=-N}^{N}C_{n_{4}}\phi(u_{i}+\xi_{41}n_{4}h,v_{j}+\xi_{42}n_{4}h,w_{k}+\xi_{43}n_{4}h)\nonumber \\
+&\sum_{n_{5}=-N}^{N}C_{n_{5}}\phi(u_{i}+\xi_{51}n_{5}h,v_{j}+\xi_{52}n_{5}h,w_{k}+\xi_{53}n_{5}h)\nonumber \\
+&\sum_{n_{6}=-N}^{N}C_{n_{6}}\phi(u_{i}+\xi_{61}n_{6}h,v_{j}+\xi_{62}n_{6}h,w_{k}+\xi_{63}n_{6}h)]\nonumber \\
+&[V_{ion}(u_{i},v_{j},w_{k})+V_{H}(u_{i},v_{j},w_{k})+V_{XC}(u_{i},v_{j},w_{k})\nonumber \\
+&V_{\theta}(u_{i},v_{j},w_{k})]\times\phi(u_{i},v_{j},w_{k}),
\end{align}  
where $C_{n_{i}}=f_{i}c_{n_{i}}$, $N$ is the order of the finite-difference expansion. The choice of $\xi$ as $-1$, 0, or 1 depends on the lattice vectors of the crystal. Given that the Laplacian operator extends to only a few neighbors around each grid point, Eq. (\ref{eq:express}) is a sparse matrix. We solve it here to obtain the minimum energy $E^{OF}_{min}[\rho]$ by implementing an energy-minimization scheme. Previous works have shown that the Truncated Newton (TN) method\cite{Nash200045} is one of the most efficient; \cite{garcia2007efficient, PROFESS1.0,hung2012preconditioners} therefore, it is employed in our approach.

\subsection{\label{Algorithm} Algorithm for Energy Minimization}
\begin{figure}
\begin{center}
\includegraphics[width=8.5cm]{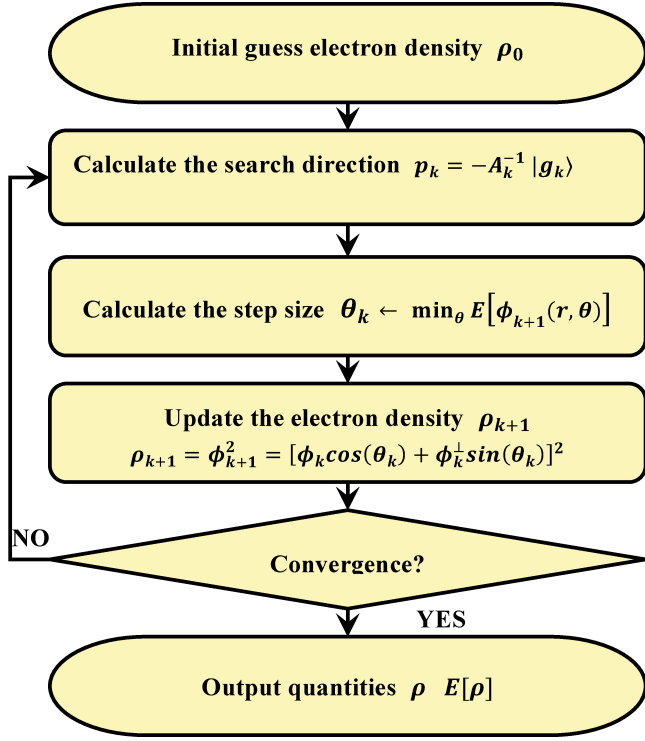}
\caption{\label{fig:flow chart}Flow chart of ATLAS.}
\end{center}
\end{figure}

Our scheme selects $\sqrt{\rho}$ as the variable to minimize the total energy. The flow chart of ATLAS code is shown in Fig. \ref{fig:flow chart}. The procedure followed  comprises three major steps. First, an initial guess of the electron density $\rho_{0}$ is required for a trial solution of $\phi^{2}$. Note that the initially guessed electron density is derived from the model of a homogeneous electron gas (the average charge density of the system).

Second, the ground-state electron density is obtained by minimizing the total energy using the TN method based on the initial guess. The TN algorithm for energy minimization consists of two iterations: an outer iteration that approximates the descent direction $p$ as the direction for minimizing energy, and an inner iteration that determines the step size $\theta$ by a line search to ensure an energy decrease.\cite{garcia2007efficient,Nash200045,PROFESS1.0,hung2012preconditioners} The details of this step include three procedures.

(i) According to the TN scheme, the search direction $|p_{k}\rangle$ at iteration $k$ is simply determined by the quantities of the current iteration. $|p_{k}\rangle$ can be written as follows: 
\begin{equation}
|p_{k}\rangle=-A_{k}^{-1}|g_{k}\rangle,
\label{eq:p}
\end{equation} 
where $|g_{k}\rangle$ is the gradient of $L_{k}(\phi)$ according to Eq. (\ref{eq:gl}), which can be written as
\begin{equation}
|g_{k}\rangle=2\left\{H_{k}|\phi_{k}\rangle-\mu_{k} |\phi_{k}\rangle\right\},
\end{equation}
with
\begin{equation} 
\mu_{k}\equiv\frac{\langle\phi_{k}|H_{k}|\phi_{k}\rangle}{N_{e}}.
\end{equation}   
$A_{k}$ is the approximate Hessian matrix of $L_{k}$:
\begin{equation}
A_{k}=\frac{\delta^{2}L_{k}}{\delta \phi(\bm{r})\delta\phi(\bm{r'})}.
\end{equation}
As in previous works,\cite{garcia2007efficient,PROFESS1.0} we rewrite Eq. (\ref{eq:p}) as a linear equation to determine $|p_k\rangle$:
\begin{equation}
A_{k}|p_{k}\rangle=-|g_{k}\rangle.
\end{equation}
This equation can be solved using the linear conjugate gradient method.\cite{hestenes1952methods} We compute $A_{k}|p\rangle$ using the first-order finite-difference approximation rather than attempt the explicit evaluation of $A_{k}$:
\begin{equation}
A_{k}|p\rangle\approx\frac{|g(\phi+\epsilon p)\rangle-|g(\phi)\rangle}{\epsilon}.
\end{equation}
This ensures that the computational cost of our approach are linear scaling.

(ii) The step size $\theta_{k}$ is determined by line search with the normalization constraint of $|\phi_{k+1}\rangle$. $|p_{k}\rangle$ is further orthogonalized to $|\phi_{k}\rangle$ and normalized to $N_{e}$.
\begin{equation}
|\phi_{k}^{'\perp}\rangle=|p_{k}\rangle-\frac{|\phi_{k}\rangle\langle\phi_{k}|}{N_{e}}|p_{k}\rangle
\end{equation}
\begin{equation}
|\phi_{k}^{\perp}\rangle=\sqrt{\frac{N_{e}}{\langle\phi_{k}^{'\perp}|\phi_{k}^{'\perp}\rangle}}  |\phi_{k}^{'\perp}\rangle
\end{equation} 
 $|\phi_{k+1}\rangle$ then is updated by
\begin{equation}
|\phi_{k+1}\rangle=|\phi_{k}\rangle \cos(\theta_{k})+|\phi_{k}^{\perp}\rangle\sin(\theta_{k}),
\end{equation} 
 where the value of $\theta_{k}$ is determined by line search\cite{brent2013algorithms,press2007numerical,gill1981practical} with the Wolfe conditions to ensure it lies toward lower energy.\cite{hung2012preconditioners}
\begin{equation}
\theta_{k}\leftarrow \underset{\theta}{min} E\left[\phi_{k+1}(\bm{r},\theta)\right]
\end{equation}

(iii) When the step size $\theta_{k}$ is determined, the new electron density can be derived by $\rho_{k+1}=\phi^{2}_{k+1}$. The process is repeated until both the gradient of Lagrange $|g\rangle$ and the variation of total energy are smaller than the given tolerances.

Finally, the third step involves calculating the total energy or other related physical quantities of a given structure from the ground-state electron density.
\section{numerical results}
We consider here three bulk systems of Mg, Al, and Al$_{3}$Mg to benchmark the above formalism (as implemented in our ATLAS code) for accuracy and computational efficiency. The calculation employs the local density approximation (LDA) for electron exchange and correlation as parametrized by Perdew and Zunger.\cite{PhysRevLett.45.566,PhysRevB.23.5048} The local pseudopotentials of Mg and Al are constructed by our OEPP scheme for their respective electronic configurations of $3s^{1}3p^{1}$ and $3s^{2}3p^{1}$. The core cutoff radii are 2.6 a.u. for Mg and 2.2 a.u. for Al.
\subsection{Tests of Real-Space OF-DFT Convergence}
\begin{figure}[!htb]
  \centering
  \subfigure[]{
    \label{fig:gap} 
    \includegraphics[width=8.5cm]{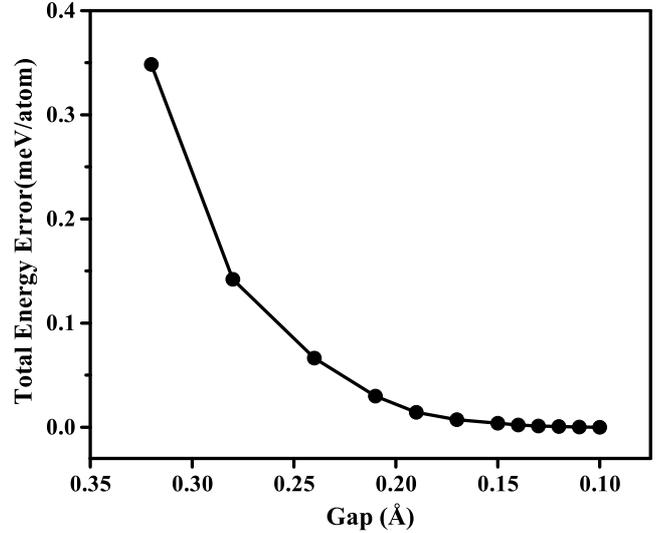}}
  \subfigure[]{
    \label{fig:order} 
    \includegraphics[width=8.5cm]{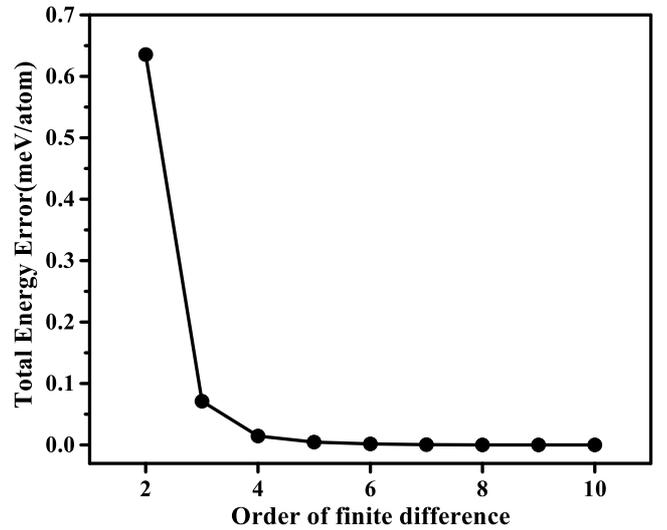}}
  \caption{ Effect of (a) grid spacing and (b) order of finite-difference approximation on the total energy of bulk Mg with a bcc lattice}
  \label{fig:converge} 
\end{figure} 
Our real-space finite-difference implementation of OF-DFT has two controllable parameters that critically influence the accuracy of the calculations: the order of finite difference expansion and the grid spacing \textit{h}. These parameters are chosen depending on the convergence test of the total energies of the systems. The grid spacing in real space is related to the plane-wave cutoff energy ($E_{cut}=\pi^{2}/2h^{2}$) in reciprocal space. Here we show in a real application how to choose the values of these parameters. We run ATLAS code on calculations of total energy for bulk Mg with a body-centered cubic (bcc) lattice. Fig. \ref{fig:converge} shows that a fourth-order finite-difference expansion and a grid spacing of 0.18 \r{A} are sufficient for a well-converged total energy (~0.1 meV/atom). Similar results are also found for bulk Al. Therefore, these two values are adopted for all the following calculations on systems of bulk Mg, Al, and Al$_{3}$Mg.

\subsection{Computational Accuracy}
For benchmarking our ATLAS program, bulk properties of hexagonal close-packed (hcp) Mg and face-centered cubic (fcc) Al and Al$_{3}$Mg are calculated and compared with those calculated by the CASTEP code\cite{segall2002first} within KS-DFT. Our ATLAS calculations employ the WGC formula for KEDF\cite{PhysRevB.60.16350} (with the following parameters: $\gamma =2.7$, $\alpha=(5+\sqrt{5})/6$, and $\beta=(5-\sqrt{5})/6$) and the OEPP local pseudopotential. Note that the WGC form of KEDF is known to describe accurately various bulk properties of Mg and Al. For meaningful comparison, the CASTEP calculations adopted the same OEPP local pseudopotentials as used in our method. The calculated equilibrium volumes, total energies, and bulk moduli are listed in Table \ref{tab:bulk}. Our ATLAS results are in excellent agreement with those obtained by CASTEP; the noticeable small differences stem from the difference between the kinetic energy terms used in the two codes.
\begin{table}[!htb]
\caption{Bulk properties obtained by OF-DFT and KSDFT methods: equilibrium volume ($V_{0}$ per atom in \r{A}$^{3}$), total energy ($E_{0}$ in eV per atom), and bulk moduli ($B_{0}$ in GPa)} 
\begin{center}
\begin{tabular}{llccc}
\hline
systems&methods & $V_{0}$ & $E_{0}$ & $B_{0}$ \\
\hline
Mg& KS  &22.023    & -24.588& 36.5\\
&OF  &22.225    &-24.577&35.0\\
\hline
Al&KS &18.029&-56.799&69.4\\
&OF &18.435&-56.801&67.4\\
\hline
Al$_{3}$Mg&KS&19.031& -48.767& 55.2\\
&OF&19.019&-48.757&57.5\\
\hline
\end{tabular}
\label{tab:bulk}
\end{center}
\end{table}
\begin{figure*}[!htb]
  \centering
  \subfigure[]{
    \label{fig:Al001} 
    \includegraphics[width=6.0cm]{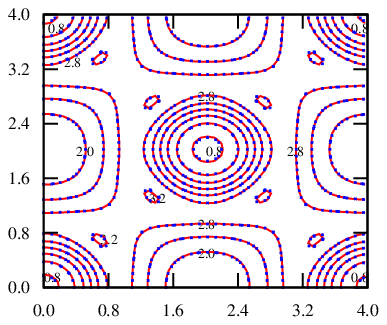}}
  \subfigure[]{
    \label{fig:Al011} 
   \includegraphics[width=6.0cm]{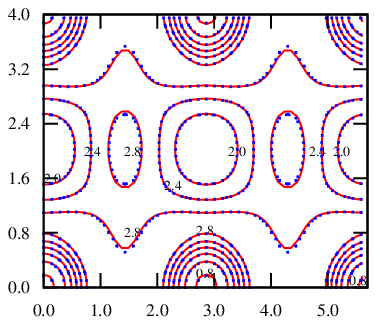}}
   \subfigure[]{
    \label{fig:Mg001} 
    \includegraphics[width=6.0cm]{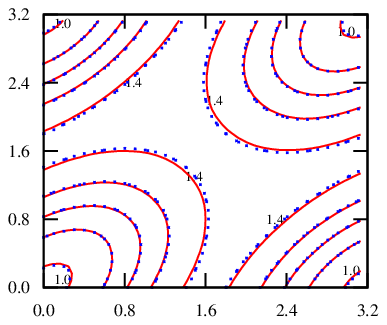}}
  \subfigure[]{
    \label{fig:Mg010} 
   \includegraphics[width=6.0cm]{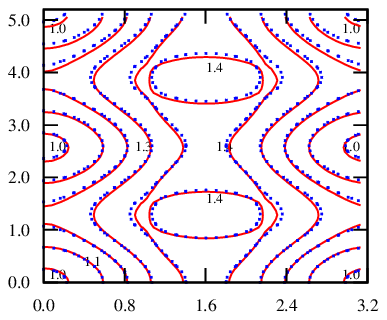}} 
 \caption{ Contour plots of electron density calculated by ATLAS (solid line) and CASTEP (dotted line). (a) The (001) and (b) (011) planes of Al; (c) the (0001)and (d) (01$\overline{1}$0) planes of Mg. Electron density and lattice vectors are given in \r{A} and (a.u./100), respectively}
\label{fig:density}
\end{figure*}

\begin{figure}[!htb]
\begin{center}
\includegraphics[width=8.5cm]{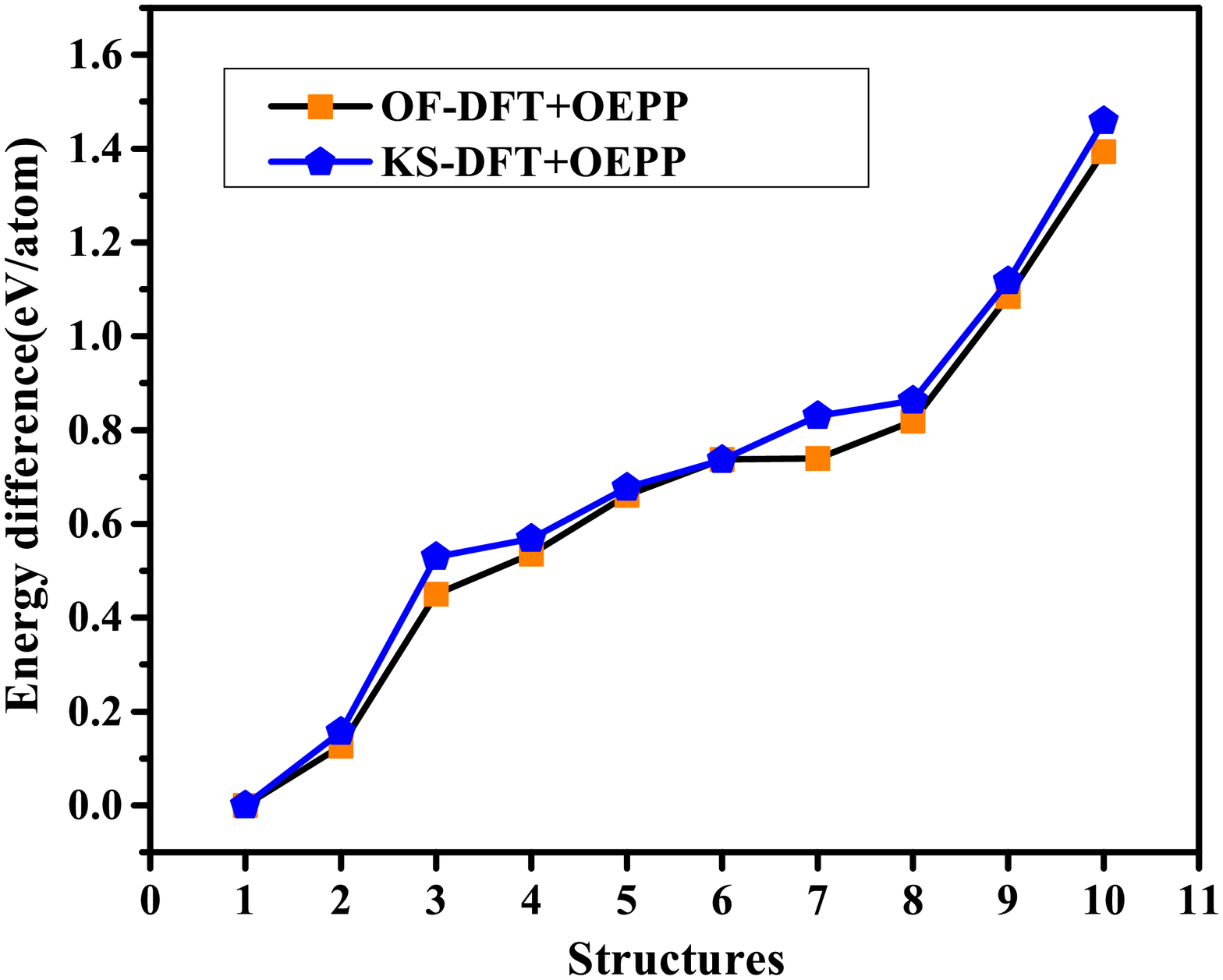}
\caption{\label{fig:pso} Relative energy differences between the first structure and another nine structures of Mg generated by CALYPSO. The numbers along the horizontal axis correspond to the structures in Table \ref{tab:pso}, which lists their detailed structural information.}
\end{center}
\end{figure}

We now focus on the fundamental quantity of electron density as calculated by the ATLAS and CASTEP codes. Fig. \ref{fig:density} shows that the two calculations give essentially identical contour plots of electron density in the (001) and (011) planes for Al and the (0001) and (01$\overline{1}$0) planes for Mg, thus supporting the accurate implementation of OF-DFT in ATLAS code.
Further validation of our method is given by randomly generating ten different structures of Mg using the CALYPSO software package \cite{calypso,wang2012} as listed in Table \ref{tab:pso}, and then calculating their total energies using both codes with the same OEPP. The results (Fig. \ref{fig:pso}) show expected small energy differences between the two sets of data, but both calculations give essentially identical structure sequences in energy order, providing further confidence in the robustness of our ATLAS code.
 \begin{table*}[!htb]
\caption{\label{tab:pso} Details of ten random structures of Mg generated by CALYPSO}
\begin{center}
\begin{tabular}{cllc}
\hline
Structure no. & Space group (Number) & Lattice parameters ($\AA$)&Wykoff position\\
\hline
1&$P2_{1}/c$ (14)&a=2.9688&4e  -0.51477   0.74802  -0.24572\\
&&b=4.7352&4e -0.78771   0.52847  -0.60354\\
&&c=11.5362&2d 0.50000  -0.00000  -0.50000\\
&&$\beta=112.3421$&\\
\hline
2&$P-6m2$ (187)&a= b=5.6880&1b 0.00000   0.00000   0.50000\\
&&c=2.6768&1d 0.33333   0.66667   0.50000\\
&&&3j 0.81547   0.18453   0.00000\\
\hline
3& $Ima2$ (46)  &a=4.2219    &4a 0.00000   0.00000   0.07444\\
&&b=8.5195&4b 0.25000   0.29658   0.01732\\
&&c=8.3406&4b 0.25000  -0.16647   0.26886\\
&&&4b  0.25000   0.55467   0.11780\\
&&&4b -0.25000   0.68084   0.26549\\
\hline
4&$P3m1$ (156)&a=b=5.6997&1a 0.00000  0.00000   0.23588\\
&&c=5.3316&1a 0.00000  -0.00000   0.65630\\
&&&1b 0.33333   0.66667   0.52999\\
&&&1c 0.66667   0.33333   0.65311\\
&&&3d 0.18246   0.81754   0.99311\\
&&&3d  0.50622   0.49377   0.25310\\
\hline
5&$Pnn2$ (34)&a=4.6491&4c 0.22362   0.16685   0.57664\\
&&b=11.0429&4c 0.70660   0.63845   0.95432\\
&&c=2.9217&2a 0.00000   0.00000   0.08609\\
\hline
6&$P4/mmm$ (123)&a=5.9328&1c  0.50000   0.50000  -0.00000\\
&&b=5.9328&1d  0.50000   0.50000   0.50000\\
&&c=4.2617&4l 0.32079   0.00000  -0.00000\\
&&&4m 0.29232   0.00000   0.50000\\
\hline
7&$P4$ (75)&a=b=5.7234&4d 0.70357   0.72197   0.85041\\
&&c=4.5792&1a 0.00000   0.00000   0.33461\\
&&&1a 0.00000   0.00000   0.85646\\
&&&2c 0.00000   0.50000   0.03534\\
&&&2c 0.00000   0.50000   0.49923\\
\hline
8&$P222_{1}$ (17)  &a=5.3893&4e 0.69544   0.72370   1.34282\\
&&b=4.7685& 2a 0.95451   0.00000   0.50000\\
&&c=5.8367&2a 0.41997   0.00000   0.50000\\
&&&2c 0.00000   0.67949   0.75000\\
\hline
9&$P2$ (3) &a=6.2958&2e -0.66003  -0.52140   1.47599\\
&&b=7.6726&2e  -0.73000  -0.98151   1.40517\\
&&c=3.1136&2e -0.15425  -0.13749   1.09614\\
&&$\beta=94.1982$&2e  -0.74070  -0.72248   0.96147\\
&&&1b  -0.00000  -0.69511   0.50000\\
&&& 1c  -0.50000  -0.34474   1.00000\\
\hline
10&$Pmna$ (53)&a=3.3393&2b  0.00000   0.00000   0.50000\\
&&b=7.5958&4h  0.50000   0.30656   0.43343\\
&&c=5.9138&4h  0.50000   0.75004   1.12286\\
\hline
\end{tabular}
\end{center}
\end{table*}
To verify the correctness of our new implementation, we calculate the total energy of Mg, Al, Al$_3$Mg, and other more complex systems (e.g., distorted structures of fcc Mg with big cells containing atomic distortions following a frozen phonon at the smallest wave vectors and with the length of the longest cell vector varying from 14.2 to 142 \r{A}) using PROFESS\cite{PROFESS1.0} with the same OEPP and KEDF (e.g., TF$\lambda$vW $\lambda$ = 1, 1/5, 1/9) as used in ATLAS. The results show that the energy difference obtained between ATLAS and PROFESS is less than 0.1 meV/atom for all these systems.

\subsection{Computational Efficiency}
\begin{figure}[!htb]
\begin{center}
 \includegraphics[width=8.5cm]{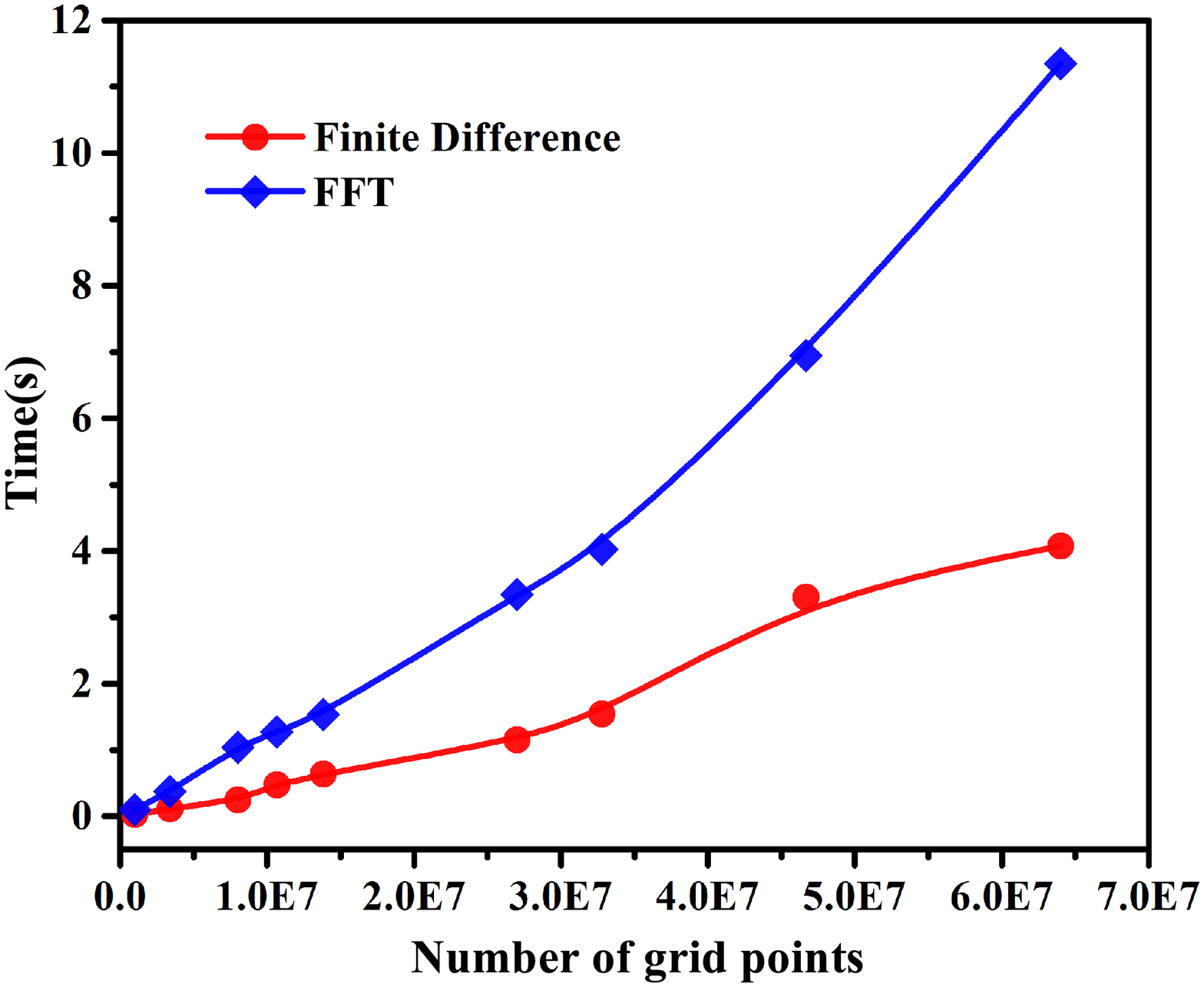}
 \caption{Timings (wall time) used to calculate vW kinetic potential in different numbers of grid points with both the finite-difference expression and the FFT-based method.} 
\label{fig:vW} 
\end{center}
\end{figure}
\begin{figure}[!htb]
\begin{center}
 \includegraphics[width=8.5cm]{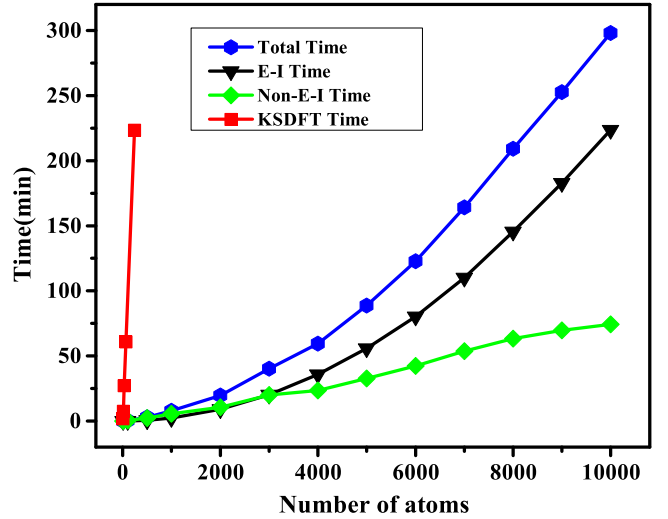}
 \caption{Timings (wall time) using ATLAS to calculate the total energy within the course of an electron-density optimization for systems of 10 to 10000 atoms in a simulated bcc Mg cell. The total time (blue line) is shown as the sum of the times for the ion--electron potential term (black line) and for all other potential terms and energy terms (green line). Also shown is the total time (wall time) cost for static energy calculation on systems of 2 to 240 atoms using CASTEP (red line)}
\label{fig:time} 
\end{center}
\end{figure}
Note that a prominent difference between our method and prior works\cite{PROFESS1.0} is that the vW term is evaluated with a finite-difference expression instead of the FFT-based approach. The (wall) times for calculating vW kinetic potential for different sizes of grid points via fourth-order finite-difference expression and the FFT-based method using FFTW\cite{frigo1998fftw} are shown in Fig. \ref{fig:vW}. The finite-difference approach is clearly computationally more efficient than the FFT method, especially for denser grid points. This is due to the different size dependence of the two methods. Assuming $N$ is the number of grid points, the computational cost of the finite-difference method is proportional to $\mathcal{O}(N)$, whereas that of FFT is proportional to $\mathcal{O}(N\log N)$. Note that our approach shows a similar advantage in dealing with the generalized gradient approximation (GGA) kinetic potentials relating to the gradient and divergence operators.

A further test of the computational efficiency of the ATLAS software package is given in the analysis of bcc Mg. Fig. \ref{fig:time} shows the (wall) times for calculations of ion--electron potential terms, all other potential terms in Eq. (\ref{eq:veff}), and the total energies within the course of an electron density optimization on systems containing 10 to 10,000 atoms using a single processor. For comparison, single-processor calculations are also performed using the DFT code of CASTEP for systems containing up to 240 atoms. Both systems use the same exchange--correlation functional and OEPP. As expected, ATLAS shows a substantial advantage in computational efficiency over KS-DFT calculation. Note that the number of iterations to reach convergence (8--10) changes little with system size. Our method therefore shows strong potential applicability to large-scale simulation.

In fact, the computational cost of ATLAS shows quadratic scaling instead of linear, because the ion--electron potential term involving an explicit treatment of structure factors scales quadratically for a periodic system.\cite{Hung2009163} To avoid the quadratic scaling problem, a particle-mesh Ewald algorithm, which has linear scaling for the ion--electron term, will be implemented in ATLAS for periodic systems\cite{Hung2009163,Essmann,PhysRevB.67.155101}. Note that for an isolated system there is no need to compute the structure factor. As a result, the ion--electron potential term will naturally show linear scaling.
\subsection{Numerical Stability}
The calculations for the Mg, Al, and Al$_{3}$Mg systems show that ATLAS is numerically stable with TF$\lambda$vW and WGC KEDF. However, previous works\cite{karasiev2012issues,PhysRevB.91.045124} have indicated that procedures implementing most GGA KEDFs are numerically unstable.  In the previous implementation, the numerical evaluation of gradient and divergence operators used the FFT. In our method, we adopted the finite-difference method to evaluate these operators for all the GGA KEDFs,\cite{garcia2007kinetic,jctc2009} and tested our method using several systems (e.g., Al and Mg with fcc, bcc, hcp, and sc  structures). The results indicate that ATLAS with GGA KEDFs is also numerically unstable except for TF$\lambda$vW, E00, and P92, which is consistent with previous work.\cite{PhysRevB.91.045124} Therefore, we believe that the numerical instabilities of most GGA KEDFs originate from unphysical electron density produced by the singular and unphysical kinetic potential\cite{karasiev2009,karasiev2012issues,PhysRevB.91.045124} during the process of optimizing the electron density.
\section{conclusion}
We developed an efficient \textit{ab initio} method for the numerical solution of OF-DFT for large-scale simulations on periodic systems, and coded it into the ATLAS software package. Our method employs the real-space finite-difference formulation and the scheme of energy minimization to yield both computational accuracy and efficiency for large-scale simulations. The performance of our method is well tested by designed static simulations for periodic systems of Mg, Al, and Al$_{3}$Mg, as well as comparison with data obtained by previous OF-DFT (PROFESS) and KS-DFT software packages (CASTEP). The results reveal that, except for the ion--electron term, the computational costs of the calculations of all other potential terms scale linearly with system size for periodic systems. Our future developments of ATLAS code will focus on the implementation of linear scaling particle mesh Ewald algorithms in an effort to achieve linear scaling on the ion--electron term,\cite{PhysRevB.67.155101,Toukmaji199673,Hung2009163} more efficient algorithms for energy minimization,  compatibility with non-periodic systems, parallel computing, and the evaluation of force and stress for ion and cell relaxations. We believe that ATLAS will become an alternative method for large-scale \textit{ab initio} simulations.
\section{acknowledgments}
The authors acknowledge funding support from the National Natural Science Foundation of China (Grant Nos 11274136 and 11404128), the Postdoctoral Science Foundation of China (Grant No. 2014M551181), the 2012 Changjiang Scholar of the Ministry of Education, and the China 973 Program (Grant No. 2011CB808204). Some of the calculations were performed at the High-Performance Computing Center of Jilin University.
\nocite{}
\bibliographystyle{apsrev4-1}
\bibliography{ATLAS}
\end{document}